# Least Effort Strategies for Cybersecurity

Sean P. Gorman*, Rajendra G. Kulkarni,
Laurie A. Schintler, Ph.D., and  Roger R. Stough, Ph.D.

School of Public Policy,
George Mason University
Fairfax, Virginia 2200, U.S.A.
*Corresponding author
e-mail: sgorman1@gmu.edu

## ABSTRACT

Cybersecurity is an issue of increasing concern since the events of September 11[th].  Many questions have been raised concerning the security of the Internet and the rest of US's information infrastructure.  This paper begins to examine the issue by analyzing the Internet's autonomous system (AS) map.  Using the AS map, malicious infections are simulated and different defense strategies are considered in a cost benefit framework.  The results show that protecting the most connected nodes provides significant gains in security and that after the small minority of the most connected nodes are protected there are diminishing returns for further protection.  Although if parts of the small minority of the most connected firm are not protected, such as non-US firms, protection levels are significantly decreased.

## INTRODUCTION

The Internet is an amalgam of thousands of interconnected networks.  Some of these networks are vast global networks like Worldcom (MCI) or Cable & Wireless while others are small local networks like a university.  The individual networks that compose the Internet are commonly called autonomous systems (AS) and number roughly 12,000 active AS's with 22,000 assigned and roughly 65,000 AS's possible (based on a 16 bit number) (1).  The task of trying to provide a minimum level of security for all these networks is a daunting effort, but one that has been increasingly highlighted as an area of importance for national security (2-4).  Innovative approaches are called for to tackle a problem of such a large scale and increasingly global nature.  Recently, researchers in the many fields have begun work concerning the fundamental structure of complex interaction of the networks that comprise the Internet (5-8).  Much of the work has revolved around the finding that the Internet at the AS and router level form a scale free network (5,9).  An understanding of the mechanics underlying the growth and evolution of the Internet provides a new perspective for the role policy can play in helping foster a more secure Internet.  A review of the literature pertaining to complex networks will be discussed with specific emphasis on implications for security in networks.  The goal of this paper is to investigate possible least effort strategies to protect the network with a minimal level of intervention.  The research will then be placed into the context of the current policy debate over cybersecurity.

## ATTACK EFFECTS AND INTERNET STRUCTURE

On Saturday January 15[th] 2003 at 5:30 UTC the SQL Slammer worm emerged from somewhere in East Asia and propagated around the globe, doubling every 8.5 seconds and infecting 90% of vulnerable machines in under 10 minutes (10). While the SQL Slammer did not carry a malicious payload, the sheer amount of traffic it produced swamped networks causing 13,000 Bank of America ATMs to become disconnected, cancelled airline flights, and disrupted elections and 911 services. The spread of SQL slammer worm was a warning of not only the speed and scope of malicious worms but the level of interdependency of the Internet with other critical infrastructures (banking and finance, transportation, medical, public safety and governance). The speed of the worm is all the more confounding when the spread and the complexity of infrastructure it traversed is considered.

The individual autonomous systems that compose the Internet broadly fall under three categories:

- *Stub AS:* It is connected to only one other AS. For routing purposes it is treated as part of the parent AS.

- *Multihomed AS:* It is connected to more than one other AS, but does not allow transit traffic. Internally generated traffic can be routed to any of the connected ASs. It is used in large corporate networks that have a number of Internet connections, but do not want to carry traffic for others.

- *Transit AS:* It is connected to more than one other AS and it can be used to carry transit traffic between other AS's. (11)

In addition to a basic typology AS's are often ranked into tiers, from 1-5. The tier 1 AS's are global networks, down to tier 5 networks consisting of local area networks for organizations and firms. The complexity of the Internet's infrastructure is daunting; these networks reside in numerous countries and fall under a wide variety of jurisdictions and most often are subject to little to no regulation, oversight or central control.

## ERROR AND ATTACK TOLERANCE OF COMPLEX NETWORKS

Scale free networks have many implications, but a far-reaching consequence of their unique structure is they are very fault tolerant but also susceptible to attack (12). Specifically, a scale free network model remains connected when up to 80% of nodes are randomly removed from the network, but when the most connected nodes are removed the average path length of the network increases rapidly, doubling its original value when the top 5% of nodes are removed (12). In short, targeting the most connected nodes can cause significant damage to a scale free network, making them highly susceptible to a coordinated and targeted attack against them. Albert et al's work was complimented by the analysis of Callaway et al (13) modeling network robustness and fragility as a percolation and by Cohen et al (14) using related methodologies. Preliminary analyses of these models on spatial network data have shown similar results when cities are the nodes and fiber connections between them are the links (15). When the most connected cities are targeted for attack the network degrades rapidly. Increased interconnection cooperation among IP transit providers reduces these effects (16). Utilizing a different model of node connectivity and path availability Grubesic et al (17) find that the disconnection of a major hub city can cause the disconnection of peripheral cities from the network. Spatial analysis of network failure has also been done for airline networks finding similar results for the Indian airline network (18).

## THE SPREAD OF VIRUSES AND WORMS IN COMPLEX NETWORKS

The scale free structure of the Internet also has implications for how viruses and worms are propagated throughout the network. Viruses and worms are not trivial computer nuisances, but high-cost problems:

By the end of August, the cost of virus attacks in 2001 totaled nearly $10.7 billion, according to researchers at Computer Economics. In previous years, computer viruses have done quite a bit of financial damage, the group says. During 2000, virus attacks cost an estimated $17.1 billion, with the Love Bug and its 50 variants doing about $8.7 billion worth of harm. And in 1999, the estimated damage was reported to be $12.1 billion…Code Red accounted for $2.6 billion in damage -- $1.5 billion in lost productivity and $1.1 billion in clean-up costs (19).

The high cost of virus and worm attacks on the Internet and connected businesses highlights the importance of understanding the nature of how these attacks spreads and what steps might be taken to mitigate them. The scale free and power law nature of the Internet illustrated by Barabasi and Albert (9) and Faloutsos et al (5) point to a methodological framework for examining the issue. Analysis of epidemics in scale free networks first reported by Pastor-Satorras and Vespignani (20), found that a wide range of scale free networks did not have an epidemic threshold[1]. The lack of an epidemic threshold meant that infections would persist and spread irrespective of the rate of the infections, however, the outcome is dependent on particular structure and topology of networks (21). This in theory could explain why viruses are rarely eradicated from the Internet and tend to spread quickly even when injected from peripheral places. Pastor-Satorras and Vespignani (27) extended this work examining immunization of complex networks including an empirical test of the Internet at the AS level. In their test a SIS (susceptible-infected-susceptible) model was implemented, where half of the nodes in the network were infected and then nodes were immunized and the effect on infection rates were recorded. They found that targeted immunizations performed significantly better than uniform immunization strategies.

---

[1] The epidemic threshold is the point at which the percentage of unvaccinated people is high enough to risk an epidemic.

Dezso and Barabasi (22) directly addresses the prospects of stopping such viruses, finding that traditional methods did not succeed is slowing spreading rates or eradicating viruses. The authors' instead found that selectively protecting the most connected nodes in the network could restore an epidemic threshold and "potentially eradicate a virus" (p.1). The study also points that a policy approach based on a "protect the hubs" strategy is cost effective, expending resources on only a few targeted nodes (22 p.3). The Dezsos and Barabasi (22) study, based on theoretical models instead of empirical data, leaves some question of how effective their strategy would be with actual networks. A recent study by Newman et al (23) studied a 16,881-user email network to determine how viruses would spread across the network. While the structure of the network was not the power law distribution seen in the theoretical scale free models discussed above, the network's exponential distribution still reacted similarly to the predicted models. Protecting the most connected email users (in the form of anti-virus software or other measures) in the network had significantly better results than randomly protecting users across the network.

The collective work on the nature of complex networks and spread of worm/virus points to a possible fruitful approach for policies that could help provide greater cybersecurity. Questions, though, still remain as to: how a "protect the hubs" strategy would play out across the Internet as whole and what level of protection would be needed to gain the maximum level of security with the minimal level of investment. Is there a distinctive phase transition where protecting a certain percentage of nodes results in a big jump in overall network security? Further, considering the global nature of the Internet can any one country implement policies that would affect enough of the network to make an appreciable impact on global network security.

## METHODOLOGY

In order to accomplish this task the problem will be simplified and examined at a macro level. The approach will be to examine the AS level topology of the Internet to determine what minimal level of protection will be required to protect the overall health of the network and prevent the wide scale spread of malicious infections. AS nodes will be selected for protection (i.e. when a worm encounters the node it will not become infected or pass along the node to others) and the selection will be tested to see how it affects the spread of an infection in the network. This approach looks at proactive measures to stop malicious infection, instead of previous research that examines containing or eliminating existing worms (27, 28). Considering the rapid spread of the SQL Slammer worm this could be a worthwhile path of investigation. From a higher level the simulation does not aim to replicate any specific malicious infection but instead illustrate the effectiveness of strategies and which networks would be the most beneficial to protect, by whatever available means. The AS's selected for protection will be first random and second based on their connectivity in the network. The threshold of AS's needed for protection will then be tested to determine at what point an acceptable level protection has been achieved. If the "protect the hubs" strategy proves a prudent strategy, further tests will be employed to determine what percentage of hubs is required for a least effort strategy to provide an adequate level of cybersecurity. Since it was not possible to acquire cost data for protection least effort is simply defined as the minimal number AS's that need to be protected. It is assumed there would be a wide variation in cost depending on the size of the AS. Further, how these AS's would be protected will not be endeavored, and the non-realistic assumption of 100% protection will first be assumed.

Each node in the network analyzed will be an individual autonomous system connected to the Internet. The data for this analysis was obtained from the University of Michigan's Internet Topology Project[2] and is based on data extracted form Oregon Route views on September 30[th], 2001 and consists of 11,955 individual autonomous systems. The AS data was then analysed utilizing two different approaches a weak worm and a strong worm. Worm, in this case, is just a generic term for a malicious infection that affects the Internet at a network level as opposed to a virus, which typically is transmitted through email. The simulation is intended to look at how infections spread from one firm's network to another and not at the IP address level that worms have used to propagate in the past. The weak worm and strong worm will both be run with a "protect the hubs strategy" with the most connected node being protected first, the next most protected second, and so on. For purposes of simplicity the protected nodes in these simulations will be referred to as the "core". The algorithm for the weak and strong worms is listed below:

## WEAK WORM ALGORITHM

1. Input: **AS** network of $n$ nodes
2. Represent $n$ nodes as a vector $\mathbf{V} = [v_1, v_2, \ldots, v_n]$.
3. Assign a value +1 to each node of $\mathbf{V}$ so as to identify these nodes as NOT INFECTED.
4. Initialize empty vectors PROTCORE **PC**, the FIFO queue POTINFECTED **PI** and reinfection counter $r$ and revisit-protected-core counter $u$ to zero.
5. Initialize immediate neighbor vectors **B** and **S**.
6. Pick a random node $v_i$ from **V** such that $v_i \notin \mathbf{PC}$.
7. Remove $v_i$ from **V** and add it to PROTCORE **PC** and assign a value of 0 to identify this node as being part of the protected core.
8. Pick a random node $v_j$ from **V**.
9. Remove node $v_j$ from **V** and put node $v_j$ in the vector INFECTED **I**
10. Assign a value of $-1$ to $v_j$
11. Find immediate neighbors **S** of $v_j$ and put them in a FIFO queue POTINFECTED **PI**.
12. While PI(1) $\neq v_j$
13. Remove first node $k = $ PI(1) from **PI**.
14. If $k \in \mathbf{PC}$, then $u = u + 1$ % Its already in the protected core
15. Else If $k \in \mathbf{I}$ then $r = r + 1$ % Its reinfection
16. Else add it to vector INFECTED **I** and assign a value of $-1$.
17. Find immediate neighbors **B** of $k$ and add them to FIFO **PI**.
18. IF **PI** $\neq$ [ ] (not empty) go to step 12.

[2] This project is supported in part by NSF Grant No. ANI-0082287, by ONR Grant No. N000140110617, and by AT&T Research.

19. Else Output: $r$, $u$, **I** and **PC** and break out of the loop beginning in step 6, else Goto setp 6.

Thus, as soon as the infection points back to the very first node that started infection, the process of infection is stopped.

## STRONG WORM

1. Input: **AS** network of $n$ nodes.
2. Compute the connectivity vector **V** for a given **AS** network.
3. Sort the vector **V** such that **V**= $[v_1, v_2, \ldots . v_n]$, where $v_1 > v_2 > \ldots > v_n$;
4. Assign a value +1 to each node of **V** so as to identify these nodes as NOT INFECTED.
5. Initialize empty vectors PROTCORE **PC**, the FIFO queue POTINFECTED **PI** and reinfection counter $r$ and revisit-protected-core counter $u$ to zero.
6. Initialize immediate neighbor vector **B** and **S**.
7. Select an arbitrary number of top m nodes from V and call this CORE **C** such that **C** = $[c_1 , c_2, \ldots . c_m]$ where $m << n$.
8. For $i$ =1 to $m$ pick node $c_i$ from **C** and put it in PROTCORE **PC** and assign a value of 0 to identify as being part of the protected core.
9. Pick a random node $v_j$ from V such that $v_j \notin$ **PC**.
10. Remove node $v_j$ from **V** and put node $v_j$ in the vector INFECTED **I**.
11. Assign a value of $-1$ to $v_j$.
12. Find immediate neighbors **S** of $v_j$ and put them in a FIFO queue POTINFECTED **PI**.
13. Remove node $k = PI(1)$ from **PI**.
14. If $k \in$ **PC**, then $u = u+1$ %% Its already in the protected core
15. Else If $k \in$ **I** then $r = r+1$ %% Its reinfection
16. Else add it to vector INFECTED **I** and assign a value of $-1$.
17. Find immediate neighbors **B** of $k$ and add them to FIFO **PI**.
18. IF **PI** $\neq$ [ ] (not empty) go to step 13. Else Output: $r$, $u$, **I** and **PC**.
19. Goto setp 4.

To provide a comparison for the "protect the hubs" strategy the strong worm algorithm will be run but the protected nodes in the core will be chosen randomly, instead of by connectivity. The random strong worm algorithm is below:

## STRONG WORM RANDOM

1. Input: **AS** network of $n$ nodes
2. Represent $n$ nodes as a vector **V** = $[v_1, v_2, \ldots . v_n]$.

3. Assign a value +1 to each node of **V** so as to identify these nodes as NOT INFECTED.

4. Initialize empty vectors PROTCORE **PC**, the FIFO queue POTINFECTED **PI** and reinfection counter $r$ and revisit-protected-core counter $u$ to zero.

5. Initialize immediate neighbor vectors **B** and **S**.

6. For a fixed number of iterations

7. Pick a random node $v_i$ from **V** such that $v_i \notin$ **PC**.

8. Remove $v_i$ from **V** and add it to PROTCORE **PC** and assign a value of 0 to identify as being part of the protected core.

9. Pick a random node $v_j$ from **V**.

10. Remove node $v_j$ from **V** and put node $v_j$ in the vector INFECTED **I**.

11. Assign a value of $-1$ to $v_j$

12. Find immediate neighbors **S** of $v_j$ and put them in a FIFO queue POTINFECTED **PI**.

13. Remove first node $k = \text{PI}(1)$ from **PI**.

14. If $k \in$ **PC**, then $u = u + 1$  %  Its already in the protected core

15. Else If $k \in$ **I** then $r = r + 1$ % Its reinfection

16. Else add it to vector INFECTED **I** and assign a value of $-1$.

17. Find immediate neighbors **B** of $k$ and add them to FIFO **PI**.

18. IF **PI** $\neq$ [ ] (not empty) go to step 12.

19. Else Output: $r$, $u$, **I** and **PC**.

20. Goto setp 5.

When the weak worm is run a node is randomly chosen and all of its neighbors are infected. Next one of those infected neighbors is randomly chosen and all of its neighbors are infected and the process is repeated until the infection refers back to the originating node. The worm takes a random walk across the network, infecting all the neighbors of each node in its walk. The strong worm on the other hand infects all of the neighbors instead of just selecting one node to follow. This allows the worm to infect all AS's in a rapid manner when no protection is in place. To manage the strong worm computationally a queue approach was used where the neighbors of the originally infected node are put into a queue and infected in turn. As the worm spreads each neighbor's neighbors are put into the queue and infected as well. This way the length of the queue, nodes to be infected, can be plotted along with the total number of infected nodes, total number of attempts to infect nodes per cycle, and the number of times the protected core is visited per cycle. A cycle is simply a single simulation run with $n$ number of nodes protected. The output produced by the simulation takes the worst-case infection scenarios from 15 iterations of each cycle. The results from the weak worm strong worm, and random weak and strong worm are presented below as Figures 1 to 4.

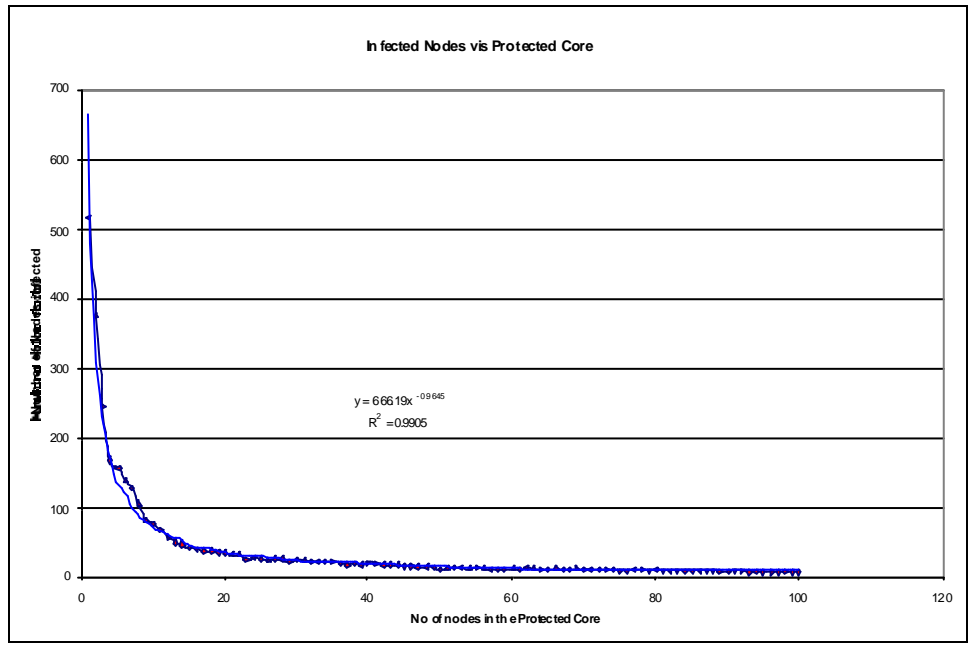

*Figure 1:* The number of nodes protected versus size of worst-case infected cluster.

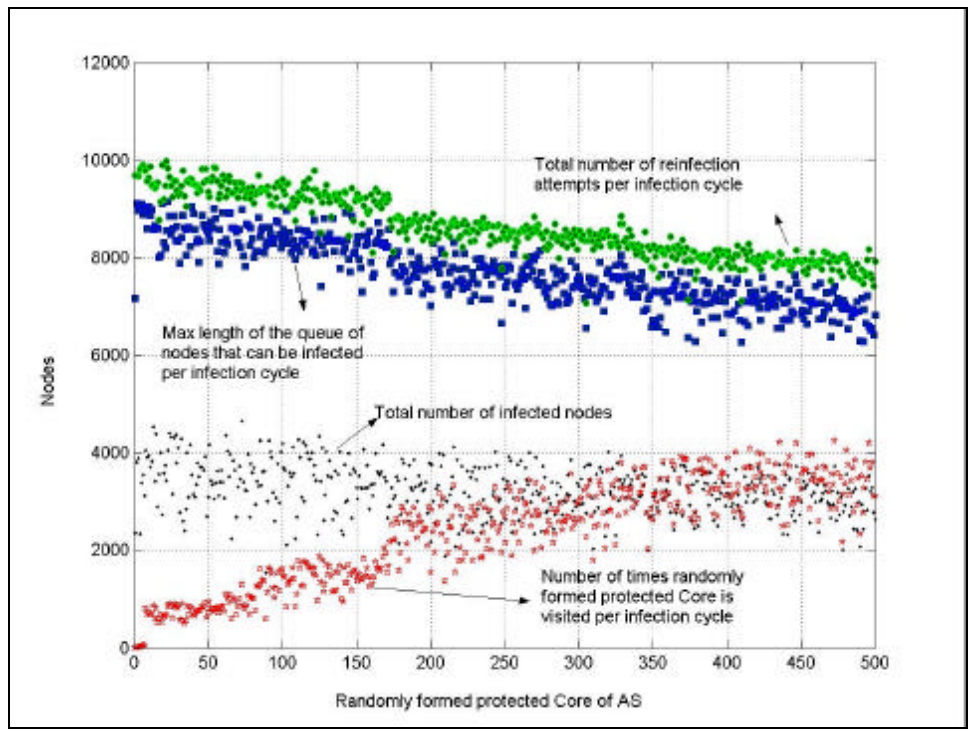

*Figure 2:* The number of nodes protected versus size of infected cluster with random protection strategy and weak worm

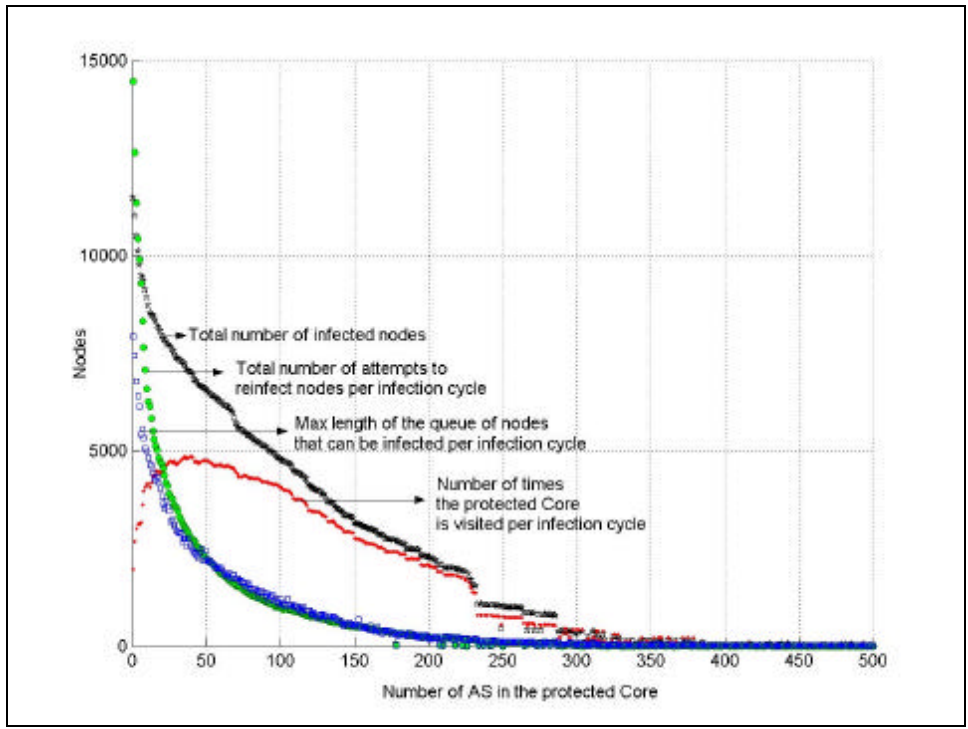

*Figure 3:* The number of nodes protected versus worst-case size of infected cluster over 15 iterations.

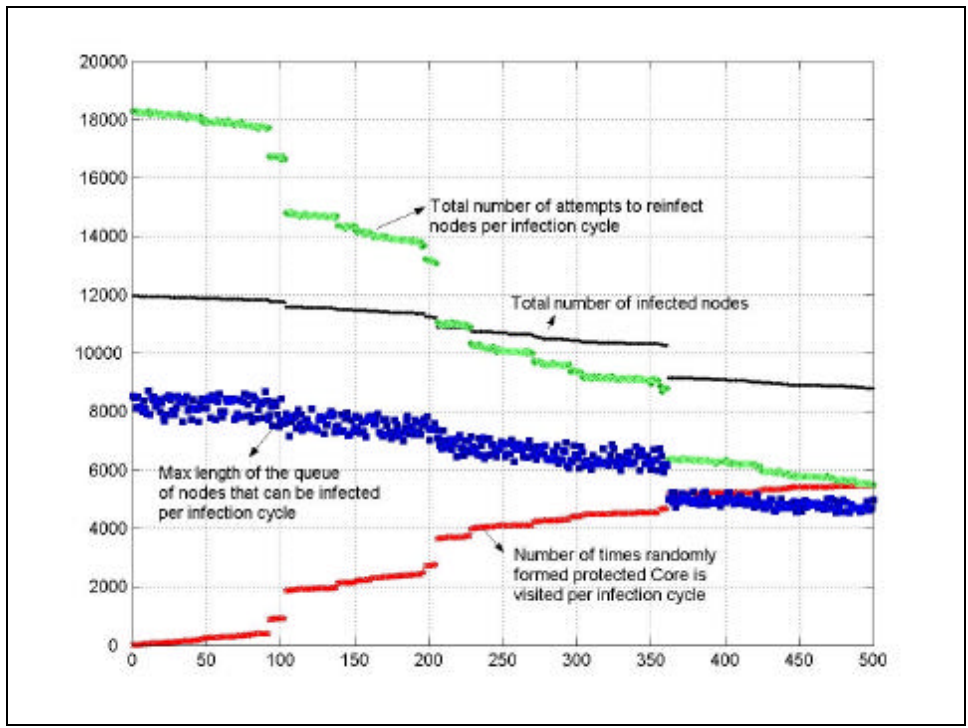

*Figure 4:* The number of nodes protected versus worst-case size of infected cluster with random protection strategy

This approach allows the testing of how increasing the size of the protected core affects the spread of virus/worm across the network of AS's. The results of the weak worm in figure 1 illustrate a precipitous drop in infection clusters with the first few nodes protected, and after about 20 nodes protected the change in infection cluster is relatively small. The sharp shift is indicative of the tight power law fit ($R^2$ = .9905) when the data is placed in a log-log format. The results seem to indicate a distinct point of inflection where there are diminishing returns for further investment in nodal protection. When this is compared to a random protection strategy with the weak worm the results are dramatic. The weak worm infections under the random strategy illustrate a random distribution with little noticeable decrease even after 500 nodes are protected.

The result of the strong worm (Figure 3) does not show the sharp power law decline seen in the weak worm, but there is still a definitive point where the infection drop off drastically (node 229) and then becomes largely ineffective (node 275). While the number of nodes requiring protection to contain the worm is larger than the weak worm, the total nodes protected are only 2.3% of the total network.

In comparison, when a random protection strategy is implemented (figure 4) little to no protection is afforded even after 500 nodes are protected. Under the random protection strategy, protecting even 500 nodes results in 8789 infected nodes, 72.3% of the total network. The one large drop in the results is because AS 701 (Worldcom) the most connected node in the network was randomly chosen.

The results of this study and particular simulation approach illustrate a significant improvement in security from a "protect the hubs strategy", although the strategy becomes less effective as a worm is made more potent. A needed extension of this work is to investigate scenarios where protection is not 100% and some worms find their way through an AS's defense. This would provide an additional level of testing for the effectiveness of the strategy.

There remains another important component of a policy perspective for this research in the composition of the top 275 AS's. How many would fall under US jurisdiction and would a US policy affecting just US firms be enough to obtain a reasonable level of security? In order to begin examining this issue the top 350 AS's registered with US addresses were compiled. Next a protection scenario was run with the most connected US firm protected followed by the next most connected through the top 350 using the strong worm methodology outlined previously. The results of the strong worm algorithm with a US only protection strategy was then plotted over the same procedure using the top 350 global AS's – the result can be seen in Figure 5.

The results illustrate similar levels of protection with the US only strategy (blue) through the top 75 AS's, then the global AS strategy (red) begins to protect more AS's than the US only strategy. After the top 100 AS's the US only strategy flat lines with over 5,000 AS's still being infected, while the global strategy continues to decrease and largely contains infections around 275 AS's protected. The result makes a preliminary case that a US only policy strategy could be an inadequate measure. International cooperation or policies that can influence foreign firm's security policies appear to be needed.

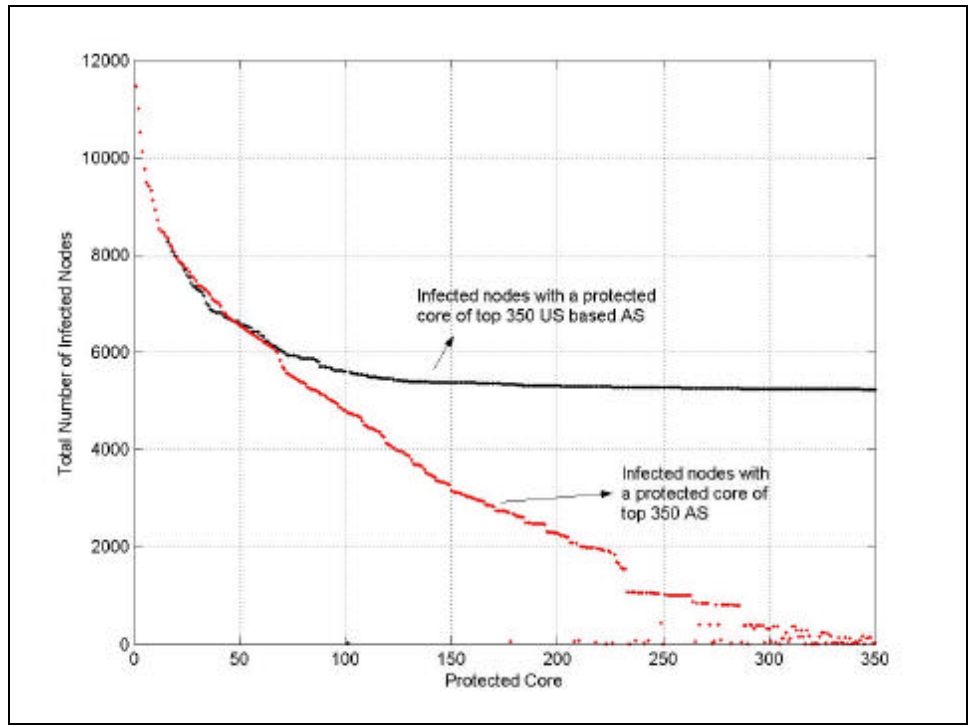

*Figure 5:* US vs. Global protection with the strong worm algorithm.

## POLICY IMPLICATIONS AND CONCLUSIONS

The result of the "protect the hubs" strategy has implications for policy in regards to best approaches to cybersecurity and critical infrastructure protection. Several studies have pointed out the fragility and vulnerability of the Internet to malicious attack (12-14). There has been debate as to the best policy approach to the current security shortcomings of the US's information infrastructure. It has been offered that there are several options for dealing with the current situation ranging from regulation, market forces, contract law, standards/best practices, insurance, or government mandated procurement requirements (24). While a full discussion of all the possible interventions for security goes outside the scope of this paper the results can shed some lights as to which general directions might bear the most fruit.

Perhaps the most persuasive argument from the results is that universal regulation is most likely an excessive approach to the problem. At the same time an uncoordinated approach fostering random protection appears to be largely ineffective. In the case of telecommunications, industry wide regulation has been most often justified in the quest to provide universal service for a population (25-26). The results of this analysis illustrates that the universal protection theoretically offered by regulation of the Internet produces minimal returns in relation to the effort to protect all the networks connected to the Internet. In fact returns diminish significantly after the protection of the top 20 nodes in the network with a weak work and the top 275 nodes with a strong worm, which constitute only .17% and 2.3% respectively of total nodes in the network.

Further questions still remain, how many firms control the top AS's? The top 20 network providers control over 70 AS's, so it is likely there are not 275 separate firms to deal with. It was not possible to perform this analysis for this paper, but it is an important extension of this research that is under investigation.

The results presented in these simulations, in term of percentages, can be deceiving, while it only requires protection of 2.3% of total nodes to obtain a high level of security the cost of protecting nodes is not equal. The most connected nodes in the network are large global networks like MCI, Sprint and AT&T. The cost of securing global networks of this size are significant and dwarf the cost of securing smaller campus networks. Needless to say protection of 2.3% of nodes would not equal 2.3% of costs.

Further, the small number of firms represented by the top 20 or top 275 AS's would seem to indicate that public – private partnerships or selective regulation to address the problem would be beneficial. The difficult task is ensuring that as many of the top AS's are protected as possible. Even with just the non-US networks removed the level of protection is significantly reduced. Also it remains to be seen if market forces or even public-private partnerships can provide adequate coverage of the top AS's. Selective regulation of the top AS's could ensure coverage but questions of equity and hampered competition and innovation could arise. Several of the alternative approaches delineated by Hunker (24) could be answers to the dilemma. For instance, if the US government has contracts with a significant number of these networks basic security requirements built into RFP's could cover a large number of firms and also provide economic incentive for compliance in the form of increased service fees to cover upgrades. Whether this selective regulation approach would provide a suitable level of cooperation among the needed core AS's and cover non-US firms remains to be seen. What is clear is coordinated and targeted security strategies provide far greater returns than random strategies. It could be possible to simulate a cooperate or defect choice with different policy scenarios through an agent based model with the data presented in this study, and this is a future direction of research.